# LTE Cellular Networks Packet Scheduling Algorithms in Downlink and Uplink Transmission: A Survey

Najem N. Sirhan, Manel Martinez-Ramon

Electrical and Computer Engineering Department,
University of New Mexico, Albuquerque, New Mexico, USA

*ABSTRACT*

*This survey paper provides a detailed explanation of Long Term Evolution (LTE) cellular network's packet scheduling algorithms in both downlink and uplink directions. It starts by explaining the difference between Orthogonal Frequency Division Multiple Access (OFDMA) that is used in downlink transmission, and Single Carrier – Frequency Division Multiple Access (SC-FDMA) is used in uplink. Then, it explains the difference between the LTE scheduling process in the donwlink and uplink through explaining the interaction between users and the scheduler. Then, it explains the most commonly used downlink and uplink scheduling algorithms through analyzing their formulas, characteristics, most suitable conditions for them to work in, and the main differences among them. This explanation covers the Max Carrier-to-Interference (C/I), Round Robin (RR), Proportional Fair (PF), Earliest Deadline First (EDF), Modified EDF-PF, Modified-Largest Weighted Delay First (M-LWDF), Exponential Proportional Fairness (EXP-PF), Token Queues Mechanism, Packet Loss Ratio (PLR), Quality Guaranteed (QG), Opportunistic Packet Loss Fair (OPLF), Low Complexity (LC), LC-Delay, PF-Delay, Maximum Throughput (MT), First Maximum Expansion (FME), and Adaptive Resource Allocation Based Packet Scheduling (ARABPS). Lastly, it provides some concluding remarks.*

## 1. INTRODUCTION

### 1.1. LTE Transmission

Long Term Evolution (LTE) uses Orthogonal Frequency Division Multiplexing (OFDM) as the basic signal format. LTE transmission in downlink is performed via the use of Orthogonal Frequency Division Multiple Access (OFDMA), while in uplink it is performed via the use of Single Carrier – Frequency Division Multiple Access (SC-FDMA) [15].

LTE uses OFDM for its high resilience to interference, its modulation format makes it very suitable for carrying high data rates, and its ability to be used in both Frequency-Division Duplex (FDD) and Time-Division Duplex (TDD) formats [7]. OFDM is higly resistant to electromagnetic interference due to the availability of multiple sub-channels, and it allows a more efficient use of the available total bandwidth as the sub-channels are closely spaced [23]. OFDM provides high data rates by transmitting a large number of multiple carriers, each transmitting a low data rate stream. Also, by properly choosing the symbol duration and carrier spacing, it is possible to efficiently modulate and demodulate the signal [8]. Therefore, LTE is an OFDMA-based technology standardized in the 3rd Generation Partnership Project (3GPP) release 8 and the following releases 9, 10, 11 and 12 to date [15].

As regards to FDD and TDD, LTE supports the use of both schemes with few differences, most distinguishable is the existence of a special frame in the case of TDD scheme. In FDD scheme, both uplink and downlink frames are transmitted on the same time slots, but over different frequencies. In TDD scheme, both uplink and downlink frames are transmitted over the same frequency, hence the existence of a special frame is needed to switch between them as Figure 1 shows [5].

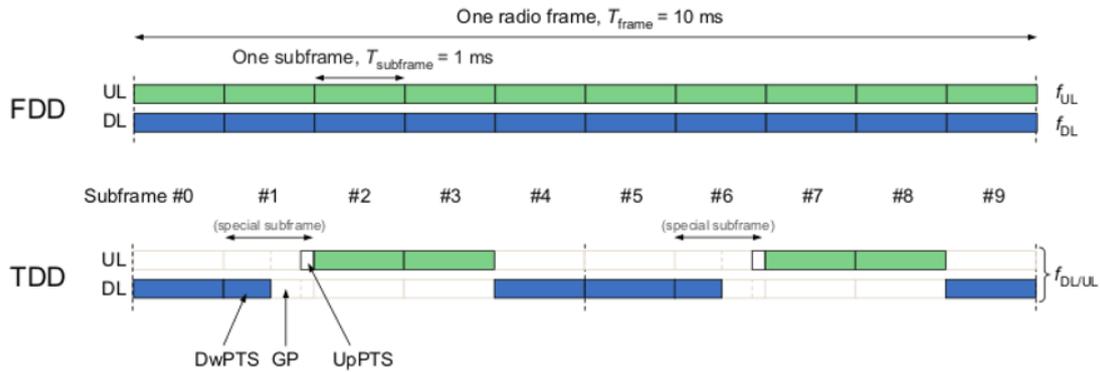

Figure 1. Uplink/Downlink time-frequency frame structure for FDD and TDD [5].

LTE transmission in downlink is performed via the use OFDMA. OFDM and OFDMA are two different variations of the same wireless broadband air interface which are often confused with each other. OFDMA is an OFDM form, which is the underlying technology [15]. The main difference between OFDM and OFDMA is based on the way they schedule users. In OFDM, the entire bandwidth of the channel is assigned to a single user for a period of time. However, in OFDMA, multiple users can share the bandwidth of the channel at the same period of time. The use of OFDMA will improve the spectrum utilization, and help in avoiding narrow-band fading and interference. A visual difference between OFDM and OFDMA is displayed in Figure 2 [2].

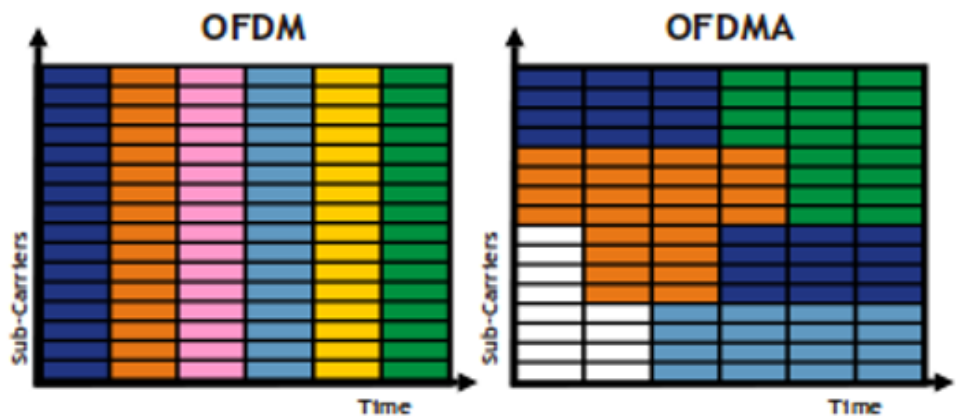

Figure 2. OFDM vs. OFDMA. Each color represents a burst of user data. In a given period, OFDMA allows users to share the available bandwidth [2].

LTE transmission in uplink is performed via the use of SC-FDMA. It is being used for its low Peak-to-Average Power Ratio (PAPR), its low-complexity while maintaining high-quality equalization in the frequency domain, its flexible bandwidth assignments, its ability to spread the information of one symbol through all the available sub-carriers "frequency diversity" to avoid a complete loss of the information modulated in the symbol in case of losing partial information on one (or even more) sub-carriers due to deep fading [22].

SC-FDMA system's performance is significantly affected by the mapping mode that is being used. Two main types of mapping modes exist that a SC-FDMA system can adopt, one is distributed, and the other is localized [22].

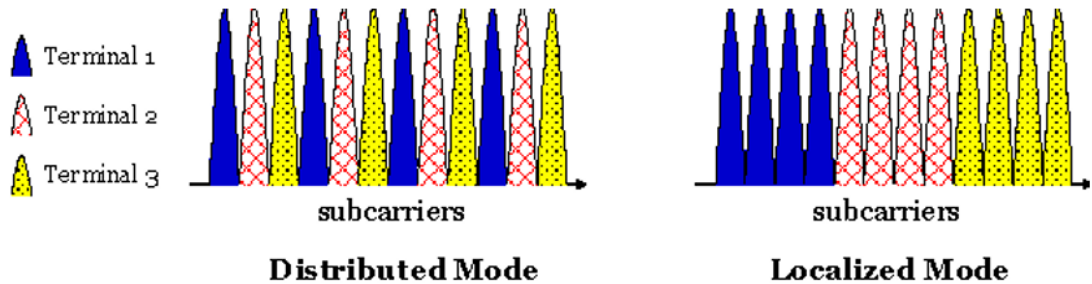

Figure 3. The difference between the distributed and the localised mapping schemes [13].

In the case of the *localized mode*, each user uses a set of adjacent sub-carriers to transmit its data, more specifically only a fraction of the total bandwidth is used by one user. The use of this mode has the advantage of achieving multi user diversity in frequency selective channel if the sub-carriers that were assigned to each user have a high channel gain. Despite this advantage, there are two drawbacks of using this mode, one is elimination of chance of getting frequency diversity in the channel, and the other one is in its requirement of Channel State Information (CSI) to map the data into the best adjacent symbols. In the case of the *distributed mode*, the sub-carriers, which are used by a user are being spread over the entire bandwidth. Spreading the information provides inherent frequency diversity. Interleaved SC-FDMA (IFDMA) is one of the common versions of distributed modes. In this mode, the assignment of sub-carriers to terminals is within an equal distance to each other, which has the disadvantage of losing user diversity [22].

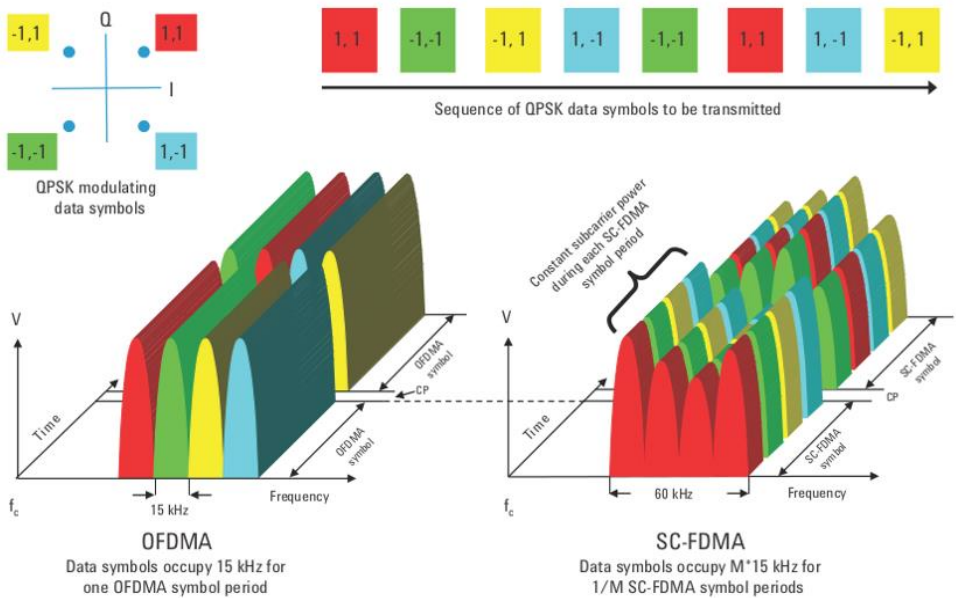

Figure 4. OFDMA vs. SC-FDMA. Where 4 sub-carriers are used for both schemes, while in real LTE signals, 12 sub-carriers are used. From the left side of the figure, each sub-carrier is colored differently to represent a different modulated Quadrature Phase Shift Keying (QPSK) symbols, that lasts for a relatively long period of time "one FDMA symbol period" when compared to it in the SC-FDMA. From the right side of the figure, a single carrier that consists of multiple sub-carriers are used, but all of these sub-carriers are modulated with the same QPSK symbol, hence the use of the same colour [9].

## 1.2. LTE Scheduling

Packet scheduling is one of the main procedures of Radio Resource Management (RRM). Packet schedulers are responsible for allocating radio resources to users' packets, and they are deployed by the Medium Access Layer (MAC) that is hosted at the eNodeB, while users' applications and connections are scheduled by the application layer [21].

In order for the radio resources to be allocated to users, a comparison that is based on a pre-defined metric has to be calculated. This metric could be viewed as the priority of each user for a specific RB, and it is performed every Transmission Time Interval (TTI) "which equals 1 ms", in order to calculate the allocation decision which is sent to the users over the Physical Downlink Control Channel (PDCCH). This means that on every TTI, the RBs should be distributed among users with the highest metric, and this sharing is done over Physical Downlink Shared Channel (PDSCH). The PDCCH contains the Downlink Control Information (DCI) messages that inform the users about the RBs, which were allocated for data transmission on the PDSCH in the downlink direction, and the RBs that were allocated to their data transmission on the Physical Uplink Control Channel (PUSCH) in the uplink direction [3].

A general model of the flow of the interaction between a downlink packet scheduler and users is displayed in Figure 5. This interaction that is repeated every TTI, represent the whole process of scheduling that can be divided into five main steps. First, the reference signal is decoded, and the Channel Quality Indicator (CQI) is computed by the user to be sent back to the eNodeB. Second, the eNodeB uses the CQI information in making the allocation decisions and filling up a RB allocating mask. Third, the best Modulation and Coding Scheme (MCS) that should be used for the data which will be transmitted to the scheduled users, will be selected by the Adaptive Modulation and Coding (AMC) module. Fourth, all the above information will be sent to the users on the PDCCH. Finally, each user reads the PDCCH and accesses to the proper PDSCH if it has been scheduled [21].

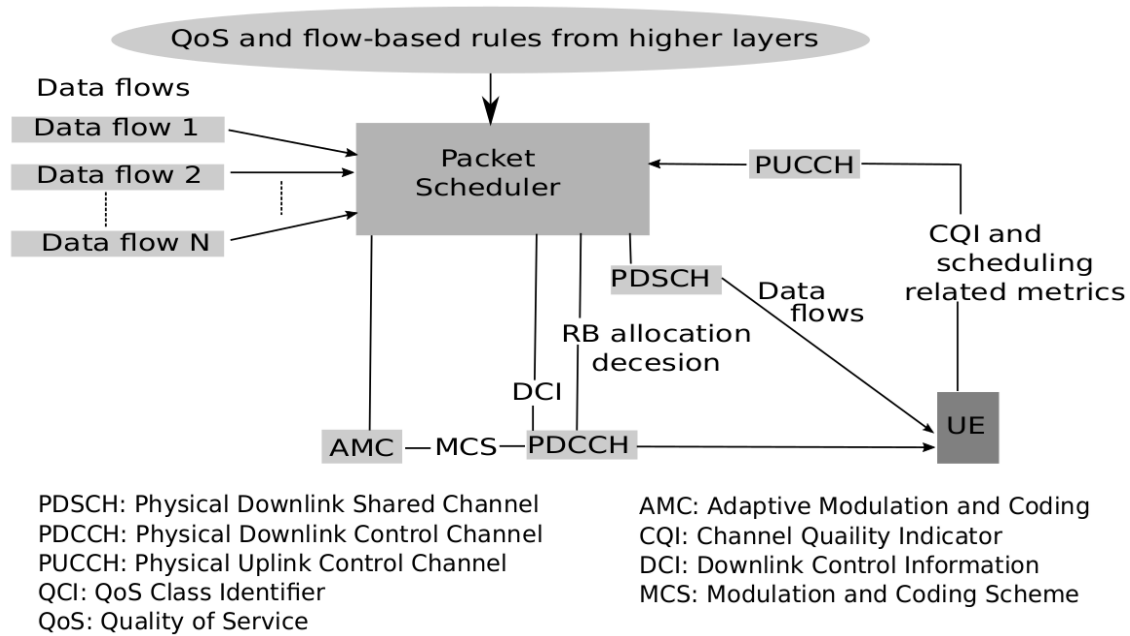

Figure 5. A general model of LTE packet scheduler [21].

The interaction between an uplink packet scheduler and users is slightly different because there is data to be uploaded by the user. This interaction can be divided into four main steps. First, the user which needs to be scheduled for uploading its data, it has to send a Scheduling Request (SR) to the eNodeB over the PUCCH format 1, which is used to carry Uplink Control Information (UCI) related to uplink scheduling. In addition to the SR, the user also sends the Buffer Status Report (BSR) and the Channel Quality Indicator (CQI). Second, when the eNodeB receives all of these information, it will decide to allocate radio resources to this user as long as it has data to transmit in the its transmission buffer, and the channel conditions are good for transmission, and there are available radio resources. Thirdly, the eNodeB will send the upllink scheduling decision back to the user over the PDCCH. Finally, the user will read the PDCCH and upload its data over the radio resources which where granted to it [17].

## 2. LTE PACKET SCHEDULING ALGORITHMS IN DOWNLINK

### 2.1. Max Carrier-to-Interference C/I

The Max C/I packet scheduling algorithm is classified as a channel-dependent scheduling algorithm. Figure 6 shows a scenario where three users are being scheduled based on their maximum C/I value, on each Transmission Time Interval (TTI), the scheduler schedule users' traffic based on the best instantaneous radio link condition. Although this algorithm optimizes the system capacity due to choosing channels with high quality, it may result in starving some users that are experiencing difficult channel's conditions such as fading [5].

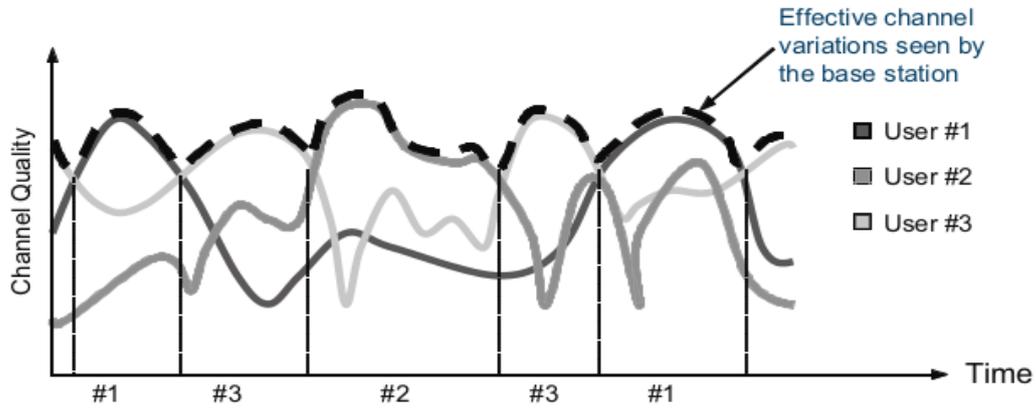

Figure 6. Channel-dependent scheduling [5]

## 2.2. Round Robin (RR) & Proportional Fair (PF)

The RR packet scheduling algorithm lets users take turns in using the shared resources, but it doesn't take into account the instantaneous channel conditions, which might lead to poor utilization of the system's capacity [5].

The PF packet scheduling algorithm schedule users' traffic in a fair way, it does this by taking into account both the experienced channel state and the past data rate when assigning radio resources to users. It aims to obtain satisfying throughput and at the same time, guarantee fairness among traffic flows. The selection is based on the following formula [18]:

$k = arg\ max\ (\ r_i(t) / R_i(t)\ )$

where $r_i(t)$ is the achievable data rate according to the instantaneous channel quality of user $i$ at $t$-th TTI, and $R_i(t)$ is the average data rate of user $i$ over a time window, and it is calculated according to the following formula [18]:

$R_i(t) = (1 - \beta) * R_i(t\text{-}1) + \beta * r_i(t\text{-}1)$

where $\beta$ is a variable ranging from 0 to 1.

A drawback of the PF scheduling algorithm is that it is only suitable to be used with non-real time traffic, and this is because it does not take into account the Quality of Service (QoS) requirements of each traffic type [18].

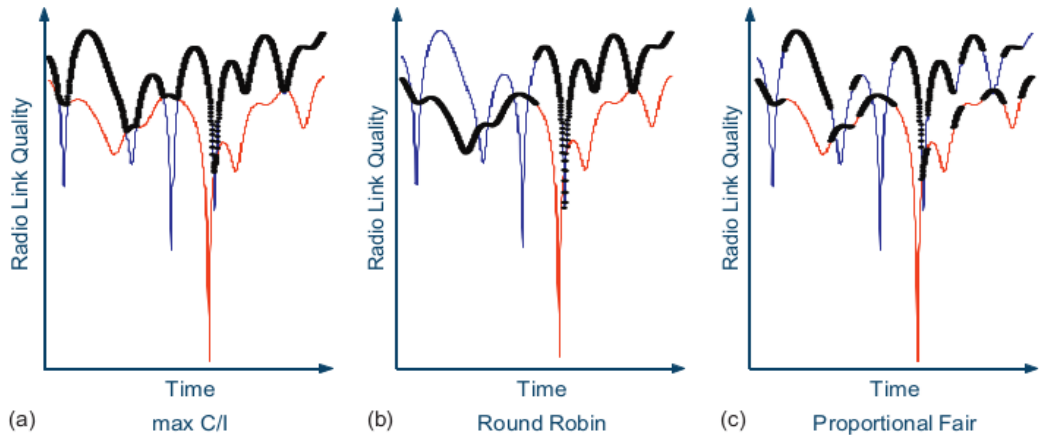

Figure 7. A comparison between the Max C/I, RR, and the PF scheduling algorithms, in a scenario of scheduling two users, the selected user is marked with black [5]

### 2.3. Earliest Deadline First (EDF)

The EDF packet scheduling algorithm is a delay restrictive algorithm. It schedules the users' traffic packets with the closest deadline expiration. A user with the closest Head of the Line (HOL) packet to the headline is chosen according to the following formula [4]:

$k = arg\ max\ (\ 1\ /\ (\ \tau_i - D_{HOL,i}\ )\ )$

where $k$ denotes the selected user with the largest metric, $\tau_i$ is the packet delay threshold of user $i$, and $D_{HOL,i}$ is the HOL packet delay of user $i$ at $t$-th TTI.

A drawback of the EDF scheduling algorithm is that it doesn't take into account the channel quality variation of wireless environments "channel-unaware", therefor it is unsuitable for use in cellular networks [4].

### 2.4. Modified EDF-PF (M-EDF-PF)

The M-EDF-PF packet scheduling algorithm combines the delay restrictive characteristics of EDF scheduling algorithm and the channel aware characteristics of the PF scheduling algorithm to ensure a good balance between throughput, fairness, and QoS provisioning. It selects a user based on the following formula [4]:

$k = arg\ max\ g_i\ (\ D_{HOL,i}\ )\ *\ (\ 1\ /\ (\ \tau_i - D_{HOL,i}\ )\ )\ *\ (\ r_i(t)\ /\ R_i(t)\ )$

where $g_i(t)$ is an increasing concave function and its curve slope varies with the change of the tunable parameters $a_i$, $b_i$, and $c_i$ and it is calculated based on the following formula:
$g_i(t) = (\ a_i * t\ )\ /\ (\ \log(1+ (b_i\ /\ t\ )) + c_i\ )$

### 2.5. Modified-Largest Weighted Delay First (M-LWDF)

The M-LWDF packet scheduling algorithm is designed to support multiple real time data users in Code Division Multiple Access – High Data Rate (CDMA-HDR) systems, it also takes into account their different QoS requirements. For example, in the case of video services, the instantaneous channel variations and delays are taken into account. It tries to balance the weighted delays of packets and to utilize the knowledge about the channel state efficiently. It choses user *j* at time *t* according to the following formula [19] [20]:

$$j = \max_i a_i \frac{\mu_i(t)}{\overline{\mu_i}} W_i(t)$$

where $\mu_i(t)$ is the data rate corresponding to user *i*'s channel state at time *t*, $\overline{\mu_i(t)}$ is the mean data rate supported by the channel, $W_i(t)$ is the HOL packet delay and $a_i > 0$, $i = 1, \ldots, N$ are weights that represent the required level of QoS.

The delay of the M-LWDF scheduling algorithm is bounded by the Largest Weighted Delay First (LWDF) scheduler. The metric that the LWDF uses are based on the system parameter that represents the acceptance probability for user *i*, in which a packet is dropped due to deadline expiration. This metric is calculated based on the following formula [19]:

$$m_{i,k}^{LWDF} = \alpha_i . D_{HOL,i}$$

where $\alpha_i = -\frac{\log(\delta_i)}{\tau_i}$

## 2.6. Exponential Proportional Fairness (EXP-PF)

The EXP-PF packet scheduling algorithm was designed specifically to support multimedia applications in an adaptive modulation and coding and time division multiplexing (ACM/TDM) systems. This means that a user's traffic can belong to a real time service or non-real time service, therefore giving the ability to assign a higher priority to non-real time services. It choses user *j* for transmission according to the following formula [10]:

$$j = \max_i a_i \frac{\mu_i(t)}{\overline{\mu_i}} \exp\left(\frac{a_i W_i(t) - \overline{aW}}{1 + \sqrt{\overline{aW}}}\right)$$

where all the parameters are the same as found in the M-LWDF, except the $\overline{aW}$ which is defined as:

$$\overline{aW} = \frac{1}{N} \sum_i a_i W_i(t)$$

If the HOL packet delays for all the users are almost the same, the exponential term is close to 1 resulting in a performance that is similar to the PF. However, if the HOL delay for one of the users becomes very large, the exponential term overrides the channel state-related terms, and the user gets a priority.

## 2.7. Token Queues Mechanism

The M-LWDF and the EXP-PF make packet scheduling decisions based on the actual packet delays. In [6], they propose to modify these algorithms by combining them with virtual tokens, this way these algorithms will not only take the delay into consideration, but also it will provide a certain minimum throughput to flows by associating a virtual token queue to each flow as shown in Figure 8.

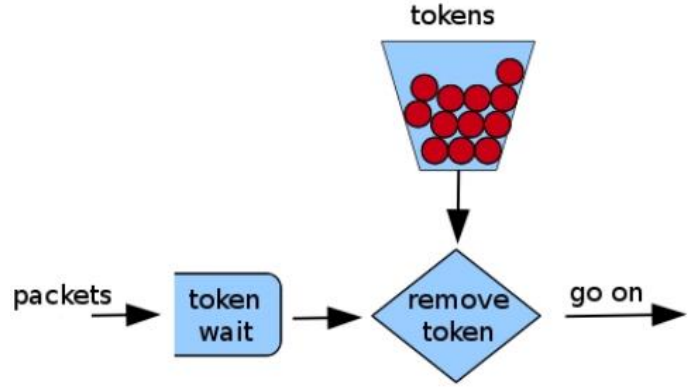

Figure 8. Token queue mechanism proposed by [10].

With this mechanism, a virtual token queue is associated to each flow, into which tokens arrive at a constant rate $r_i$, the desired guaranteed minimum throughput of flow $i$. The delay of the head of line token in the flow $i$ token queue is defined as $V_i(t)$, and it is calculated according to the following formula:

$$V_i(t) = Q_i(t) / r_i(t)$$

where $Q_i(t)$ is a counter value at time $t$ that defines the token queue length, then they used the M-LWDF and EXP-PF formulas with $W_i(t)$ being replaced by $V_i(t)$ as follows:

$$j = \max_i a_i \frac{\mu_i(t)}{\overline{\mu_i}} V_i(t)$$

$$j = \max_i a_i \frac{\mu_i(t)}{\overline{\mu_i}} \exp\left(\frac{a_i V_i(t) - \overline{aW}}{1+\sqrt{\overline{aW}}}\right)$$

After the service of a real queue, the number of tokens in the corresponding token queue is reduced by the actual amount of data transmitted.

## 2.8. Packet Loss Ratio (PLR)

The PLR packet scheduling algorithm is proposed by [12], it is suitable for real-time traffic because it calculates a priority for every user on every sub-carrier to assign the sub-carrier to the user with the highest priority. The priority function on sub-carrier $n$ is defined as follows:

$$\mu_{i,n}(t) = \begin{cases} W_i(t) \frac{r_{i,n}(t)}{\overline{r_i}} \frac{PLR_{max}^i}{PLR_i(t)}, & \text{if } PLR_i(t) > PLR_{max}^i \\ W_i(t) \frac{r_{i,n}(t)}{\overline{r_i}} \frac{k}{PLR_{max}^i}, & \text{if } PLR_i(t) < k \leq PLR_{max}^i \\ W_i(t) \frac{r_{i,n}(t)}{\overline{r_i}} \frac{PLR_i(t)}{PLR_{max}^i}, & \text{otherwise} \end{cases}$$

where $r_{i,n}(t)$ represent the channel status, $W_i(t)$ is the packet delay, $PLR_i(t)$ is the user packet loss ratio, $r_i$ is the user average throughput in the past, $PLR^i_{max}$ is the maximum packet loss ratio defined by the user, and $k$ is a non-zero constant which is far smaller than $PLR^i_{max}$.

## 2.9. Quality Guaranteed (QG)

The QG packet scheduling algorithm is designed based on the PLR scheduling algorithm to calculate the user's priority. It follows the following three rules. Firstly, if the value of the user's PLR decreases the value of the QG is less than that of the proportional fairness. Secondly, if the user's PLR approaches the maximum PLR value, the priority increases fast. Finally, if the value of the user's PLR exceeds the maximum value of the PLR, the priority is decreased to avoid wasting of the limited radio resources from the user with bad channel condition. The mathematical formula for the user priority is as follows [12]:

$$\mu_{i,n}(t) = \frac{r_i(t)}{\bar{r}_i} * f(PLR_i, W_i(t))$$

$$f(PLR_i, W_i(t)) = \begin{cases} 10^{W_i(t)/W_{max}} * 10^{PLR_i/PLR_{max}}, & \text{if } PLR_{max} \geq PLR_i \\ 10^{W_i(t)/W_{max}} * 10^{(2PLR_{max}-PLR_i)/PLR_{max}}, & \text{if } PLR_{max} < PLR_i \end{cases}$$

In [12], their simulations were applied on QG, M-LWDF, and the PLR in order to compare their performances while increasing the number of users per cell. They found that the three algorithms almost had similar performance when the system load was low, and the radio resources were enough to offer service for video services. However, when the system load increased the performance gap became larger. M-LWDF had the best performance since it maximized the packet waiting time and thus lead to a lower drop ratio.

## 2.10. Opportunistic Packet Loss Fair (OPLF)

The OPLF packet scheduling algorithm is based on calculating a simple dynamic priority function which depends on the following parameters; the HoL packet delay, the packet loss rate (PLR), and the achievable instantaneous downlink rate of each user. The HoL packet delay is calculated by assigning a buffer for each user at the eNodeB in which the arrival time for each packet could be stamped in order to be queued in a First In, First out (FIFO) order. The average packet delay corresponds to the average amount of time packets reside in the buffer. The priority function of user *i* at scheduling epoch *t* is calculated as follows [11]:

$$\text{PRF}_i(t) = \frac{R_i(t)W_i(t)PLR_i(t)}{D_{max}PLR_{thr_i}}$$

where $PLR_{thri}$ is the maximum PLR tolerated for user *i*, $W_i(t)$ is the HOL packet delay of user *i* at a current scheduling time *t*, $t_{enter}(i)$ is the time at which user *i*'s packet enters the eNodeB's buffer and is time stamped by the buffer manager, and is calculated as $W_i(t) = t - t_{enter}(i)$. $PLR_i(t)$ is the packet loss rate of user *i* at instant *t* calculated over the moving average transmission window $t_w$, and is calculated as follows:

$$\text{PLR}_i(t) = \frac{\sum_{n=t-t_w}^{n=t} dropped_i(n)}{\sum_{n=t-t_w}^{n=t} transmitted_i(n)}$$

where $transmitted_i$ and $dropped_i$ are the number of the transmitted and dropped packets for user *i*. $R_i(t)$ is the instantaneous rate of user *i* averaged over all unallocated Physical Resource Blocks (PRBs), and it is calculated as follows:

$$R_i(t) = \frac{1}{|\Phi_{URB}(t,k)|} \sum_{\varphi \in \Phi_{URB}(t,k)} r_i(t,\varphi)$$

where $\Phi_{URB}(t,k)$ denotes the set of unallocated PRBs during iteration *k* at scheduling instant *t,* and the $|\Phi_{URB}(t,k)|$ its cardinality, $r_i(t, \varphi)$ is the instantaneous rate of user *i* with PRB φ.

## 2.11. Two-level Downlink Scheduler (Upper Level Scheduler: Frame Level Scheduler (FLS), Lower Level Scheduler: PF)

The two-level downlink scheduler was proposed by [14] as shown in Figure 9, the upper level, named FLS defines on the long run how much data should be transmitted by each data source. The lowest level scheduler, named the PF allocates resource blocks in each TTI to achieve a trade-off between fairness and system throughput. Since the FLS does not take into account the channel status, the lowest level scheduler assigns RBs first to flows hosted by User Equipments (UEs) experiencing the best channel quality and then it considers the remaining ones. In particular, the lowest level scheduler decides the number of TTIs/RBs (and their position in the time/frequency domains) in which each real-time source will actually transmit its packets.

As displayed in Figure 9, the two-level scheduler has two layers, the upper level layer and the lower level layer. The upper level uses FLS scheduler, which evaluates the transmission needs of all queues at the beginning of each LTE frame. Based on the fact that the FLS has a filter finite pulse responses in which the length of the pulse is known, hence the FLS is able to grant bounded delays. The lower level layer of this scheduler uses PF which achieves a high level of fairness among multimedia flows. It assigns RBs to downlink connections belonging to UEs with the best instantaneous available data rate over the average data rate.

According to [14], in their simulation results which considered four parameters, the number of UEs, the speed, and the target delay imposed to real-time flows, and also the inter-cell interference, their proposed approach for real-time video flows outperforms the Exponential (EXP) rule, Logarithmic (LOG) rule, and Frame Level Scheduler (FLS). Their simulations were performed on an N-active traffic flows in which they all share the same wireless channel. Each traffic flow was assigned to a unique queue. VoIP and video flows were tested while users are traveling at a speed of 120Km/h with a 40ms delay bound. They compared The LOG rule, EXP rule, and FLS, and they found that the FLS allocation scheme had the best ability to provide a perceived video quality.

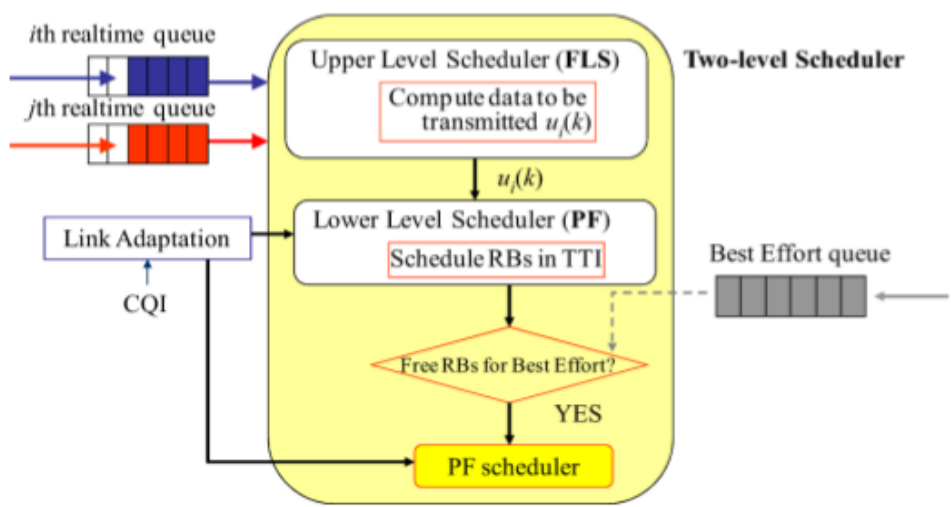

Figure 9. Two-level scheduler proposed by [14].

## 3. LTE Packet Scheduling Algorithms in Uplink

There are different LTE packet scheduling algorithms in uplink, but in most of these algorithms, the scheduler takes a matrix as an input, in which this input matrix is used for the radio resource allocation. The performance of these algorithms is affected by the method that is used to build up the input matrix. Two main methods are used in building up the input matrix; the channel dependent method, and proportional fairness method. In the case of Channel Dependent method, allocating the UE resources is considered based on the channel quality of each UE on each resource block. In order for the eNodeB to know about the channel quality of UEs on every resource block, two signals are being used in this process; one is the Sounding Reference Signal (SRS), and the other one is the Buffer Status Report (BSR). The SRS signal is being transmitted by the UE every 1ms, and from this signal the eNodeB extracts the Channel State Information (CSI) and passes it to the CSI manager that uses it to build the channel condition matrix. The BSR is also being sent by the UE to the eNodeB to indicate the amount of buffered data and their priorities. In the case of Proportional Fairness method, fairness is proportional to the channel conditions. This will guarantee that users with low channel conditions will get some of the available radio resources, but with lower amount than users with better channel conditions' priorities [16].

### 3.1. Low Complexity (LC)

Two low complexity algorithms were proposed by [24], both allocates resource block $n$ to user $k$ in a way to maximize the following difference:

$$\Lambda_{n,k} = U(R_k|I_{RB,k} \cup \{n\}) - U(R_k|I_{RB,k})$$

where the marginal utility, $\Lambda_{n,k}$ represents the gain in the utility function $U$ when resource block $n$ is allocated to user $k$, compared to the utility of user $k$ before the allocation of $n$. However, they differ in their linear search criteria. One allocates radio resource block after performing a linear search on both the users and resource blocks, it does this linear search to find the user-resource block pair that maximizes the marginal utility. The other one allocates each resource block after performing a linear search for only the users in order to find the user that maximizes the marginal utility.

### 3.2. LC-Delay

The LC-Delay packet scheduling algorithm was proposed by [6], it takes into account the end-to-end delay constraint in addition to the channel contiguity constraint. It does not only maximizes the LC scheduling algorithm's marginal utility, but it also satisfies the maximum allowed delay and guarantees a minimum throughput for each user.

It works as follows, If the maximum delay and minimum throughput requirements are satisfied for all users taking into account the adjacency resource block constraints, the LC-Delay can assign each resource block to each user in order. If not, it assigns the available and adjacent resource blocks for the user with the critical delay or throughput constraints. The drawback in this scheduling approach in the fact that some of the users might not be assigned any resource blocks.

### 3.3. PF-Delay

The PF-Delay packet scheduling algorithm was also proposed by [6], it takes into account the end-to-end delay constraint in addition to the channel contiguity constraint. It allocates a resource block that maximizes a certain metric value in a way that users will never experience a delay greater than the maximum allowed delay.

It differs from the LC-Delay in the use of the proportion between the current throughput to the total throughput instead of using the marginal utility. Also, it dose assign the resource blocks in order, but with respect to the users with the most critical delay requirement taking into account that the user has a reasonable utility value.

Neither the LC-Delay, nor the PF-Delay are efficient scheduling algorithms for the case when the number of users is smaller than the number of resource blocks, as they tend to assign one block per user, except for the last user who is assigned a large number of blocks.

### 3.4. Maximum Throughput (MT)

In the MT scheduling algorithm, maximizing the overall throughput is the main goal, and it accomplishes this goal by continuously assigning radio resources to users that are capable of transmitting data over the current TTI, which requires the user to have a high CQI value. Hence, the MT scheduling algorithm focuses mainly on the value of CQI in serving users. The users with the highest value of CQI will be served first with the required radio resources, which will lead to cell throughput enhancement. A drawback of this approach, in its lack of fairness when it comes to distributing the resources over the users with poor CQI values, such as users that exists on the cell borders "cell-edge users" [1].

### 3.5. First Maximum Expansion (FME)

In the FME scheduling algorithm, both maximizing the throughput and fairness are of concern. The FME works as follows; it starts by assigning radio resources to the user with the best channel conditions, then it expands its search in both time and frequency domains as long as the channel maintains its best condition among other users. Once another user with better channel condition is found, the FME will stop assigning resources to the first user and move into serving this user [1].

### 3.6. Adaptive Resource Allocation Based Packet Scheduling (ARABPS)

The ARABPS scheduling algorithm was proposed by [17]. Their scheduling algorithm does the radio resource allocation in two main steps; firstly, the Time Division Packet Scheduling (TDPS) that runs on all users, secondly, the Frequency Division Packet Scheduling (FDPS) that runs on the selected users out of the first step.

The TDPS chooses which users are ready to be scheduled by the eNodeB. The basis of this selection is to establish a balance between user fairness and quality of service. The selection will be determined based on average user throughput in moving windows in the time domain. And to maintain fairness between all users, the user, with low accumulated throughput in the time window will be assigned a higher scheduling priority in the next TTI.

The FDPS perform the actual frequency allocation for the users which were chosen by TDPS. The allocation of frequencies in the FDPS is based on the requirements of the chosen users and the channel conditions. Fially, it outputs the information that is related to both the time and frequency resource allocation in addition to the corresponding modulation and coding scheme parameter to each user.

In [17], they compared their scheduling algorithm with the Max C/I and RR scheduling algorithms. According to their simulation results, their algorithm made a higher throughput gain of 35% to 70% than the RR algorithm, and provided better fairness than the Max C/I algorithm.

## 4. CONCLUSION

LTE uses OFDMA in downlink transmission to surmount the effect of multipath fading problem which exists in the universal mobile telecommunications service that preceded the LTE. OFDMA has a high peak-to-average power ratio, which makes it less suitable to be used in uplink transmission. So, in order to reduce the peak-to-average power ratio, increase the efficiency of the power amplifier, and to save battery life at the user equipments, LTE uses a different access mode for uplink transmission, which is SC-FDMA.

The use of some of the downlink scheduling algorithms comes with drawbacks, for example, the use of the Max C/I may result in starving some users that are experiencing difficult channel's conditions such as fading, the use of the RR may lead to poor utilization of the system's capacity because it doesn't take into account the instantaneous channel conditions, the use of the PF is only suitable with non-real time traffic because it does not take into account the QoS requirements of each traffic type, the use of the EDF is unsuitable for cellular networks because it doesn't take into account the channel quality variation of wireless environments "channel-unaware". However, the use of other downlink scheduling algorithms can overcome these issues, for example, the use of the M-EDF-PF has an advantage of combining the delay restrictive characteristics of EDF scheduling algorithm and the channel aware characteristics of the PF to ensure a good balance between throughput, fairness, and QoS provisioning.

As regards to the use of uplink scheduling algorithms, the LC-Delay and the PF-Delay takes into account the end-to-end delay constraint in addition to the channel contiguity constraint. The use of the MT maximizes the overall throughput, but it lacks of fairness when it comes to distributing the resources over the users at the cell borders "cell-edge" users. The FME expands its search in both time and frequency domains to ensure no user gets all of the throughput even if it has a high CQI value.

**CONFLICTS OF INTEREST**

The authors declare no conflict of interest.

**REFERENCES**

[1] Ahmed, R. E., & AlMuhallabi, H. M. (2016, March). Throughput-fairness tradeoff in LTE uplink scheduling algorithms. In *2016 International Conference on Industrial Informatics and Computer Systems (CIICS)* (pp. 1-4). IEEE.
[2] Anritsu, "LTE Resource Guide." [Online]. Available: https://www.cs.columbia.edu/~hgs//teaching/ais/hw/anritsu.pdf
[3] Capozzi, F., Piro, G., Grieco, L. A., Boggia, G., & Camarda, P. (2012). Downlink packet scheduling in LTE cellular networks: Key design issues and a survey. *IEEE communications surveys & tutorials*, *15*(2), 678-700.
[4] Chandrasekhar, V., Andrews, J. G., & Gatherer, A. (2008). Femtocell networks: a survey. *IEEE Communications magazine*, *46*(9), 59-67.
[5] Dahlman, E., Parkvall, S., & Skold, J. (2013). *4G: LTE/LTE-advanced for mobile broadband*. Academic press.
[6] Delgado, O., & Jaumard, B. (2010, May). Scheduling and resource allocation in LTE uplink with a delay requirement. In *2010 8th Annual Communication Networks and Services Research Conference* (pp. 268-275). IEEE.

[7]   Electronics Notes, "OFDMA SC-FDMA & Modulation." [Online]. Available: https://www.electronics-notes.com/articles/connectivity/4g-lte-long-term-evolution/ofdm-ofdma-scfdma-modulation.php
[8]   GB, U., & Swamy, M. S. OFDM System for High Data Rate and High Mobility.
[9]   Hatoum, R., Hatoum, A., Ghaith, A., & Pujolle, G. (2014, September). Qos-based joint resource allocation with link adaptation for SC-FDMA uplink in heterogeneous networks. In *Proceedings of the 12th ACM international symposium on Mobility management and wireless access* (pp. 59-66).
[10]  Iturralde, M., Yahiya, T. A., Wei, A., & Beylot, A. L. (2011, September). Performance study of multimedia services using virtual token mechanism for resource allocation in LTE networks. In *2011 IEEE vehicular technology conference (VTC Fall)* (pp. 1-5). IEEE.
[11]  Khan, N., Martini, M. G., Bharucha, Z., & Auer, G. (2012, April). Opportunistic packet loss fair scheduling for delay-sensitive applications over LTE systems. In *2012 Ieee Wireless Communications And Networking Conference (Wcnc)* (pp. 1456-1461). IEEE.
[12]  Liu, G., Zhang, J., Zhou, B., & Wang, W. (2007, September). Scheduling performance of real time service in multiuser OFDM system. In *2007 International Conference on Wireless Communications, Networking and Mobile Computing* (pp. 504-507). IEEE.
[13]  Myung, H. G., Oh, K., Lim, J., & Goodman, D. J. (2008, March). Channel-dependent scheduling of an uplink SC-FDMA system with imperfect channel information. In *2008 IEEE Wireless Communications and Networking Conference* (pp. 1860-1864). IEEE.
[14]  Piro, G., Grieco, L. A., Boggia, G., Fortuna, R., & Camarda, P. (2011). Two-level downlink scheduling for real-time multimedia services in LTE networks. *IEEE Transactions on Multimedia*, *13*(5), 1052-1065.
[15]  Remy, J. G., & Letamendia, C. (2014). *LTE standards*. John Wiley & Sons.
[16]  Safa, H., & Tohme, K. (2012, April). LTE uplink scheduling algorithms: Performance and challenges. In *2012 19th International Conference on Telecommunications (ICT)* (pp. 1-6). IEEE.
[17]  Sha, X., Sun, J., Wu, T., & Liu, L. (2012, October). Adaptive resource allocation based packet scheduling for LTE uplink. In *2012 IEEE 11th International Conference on Signal Processing* (Vol. 2, pp. 1473-1476). IEEE.
[18]  Sirhan, N. N., Heileman, G. L., Lamb, C. C., & Piro-Rael, R. (2015). Qos-based performance evaluation of channel-aware/qos-aware scheduling algorithms for video-applications over lte/lte-a. *Computer Science & Information Technology (CS & IT)*, *5*(7), 49-65.
[19]  Sirhan, N. N., Heileman, G. L., & Lamb, C. C. (2015). Traffic offloading impact on the performance of channel-aware/qos-aware scheduling algorithms for video-applications over lte-a hetnets using carrier aggregation. *International Journal of Computer Networks & Communications (IJCNC)*, *7*(3), 75-90.
[20]  Sirhan, N. N., Martínez-Ramón, M., Heileman, G. L., Ghani, N., & Lamb, C. C. (2016, July). Qos performance evaluation of disjoint queue scheduler for video-applications over lte-a hetnets. In *Proceedings of the 7th International Conference on Computing Communication and Networking Technologies* (pp. 1-7).
[21]  Sirhan, N. N. (2017). Packet Scheduling Algorithms in LTE/LTE-A cellular Networks: Multi-agent Q-learning Approach.
[22]  Surgiewicz, R., Strom, N., Ahmed, A., & Ai, Y. (2014). LTE uplink transmission scheme.
[23]  TechTarget, "orthogonal frequency-division multiplexing (OFDM)." [Online]. Available: https://www.techtarget.com/searchnetworking/definition/orthogonal-frequency-division-multiplexing
[24]  Yaacoub, E., Al-Asadi, H., & Dawy, Z. (2009, July). Low complexity scheduling algorithms for the LTE uplink. In *2009 IEEE Symposium on Computers and Communications* (pp. 266-270). IEEE.